\documentclass[useAMS,usegraphicx,usenatbib]{mn2e}

\title[Revisiting Numerical Models of $\eta$ Car nebulae]{Revisiting
 2D Numerical Models for the 19th century outbursts of $\eta$ Carinae}

\author[R.F. Gonz\'alez et al.]{R.F. Gonz\'alez$^{1}$\thanks{E-mail:
rf.gonzalez@astrosmo.unam.mx}, A.M. Villa$^1$,
G.C. G\'omez$^{1}$, E.M. de Gouveia Dal Pino$^2$,
\newauthor A.C. Raga$^3$, J. Cant\'o$^4$, P.F. Vel\'azquez$^3$,
E. de la Fuente$^5$\\
$^{1}$Centro de Radioastronom\'ia y Astrof\'isica (UNAM),
 Ap.Postal 3-72, C.P.: 58190, Morelia, Michoac\'an, M\'exico\\
$^{2}$Instituto Astron\^omico e Geof\'isico (USP), R. do Mat\~ao
1226, 05508-090 S\~ao Paulo, SP, Brasil\\
$^{3}$Instituto de Ciencias Nucleares (UNAM),
            Ap.Postal 70-543, CP: 04510, M\'exico D.F.,M\'exico\\
$^{4}$Instituto de Astronom\'\i a (UNAM),
            Ap.Postal 70-264, CP: 04510, M\'exico D.F., M\'exico\\
$^{5}$Departamento de F\'isica, CUCEI, Universidad de Guadalajara,
 Avenida Revoluci\'on SN, Guadalajara, Jalisco, M\'exico}

\date{Accepted ---.
      Received ---;}

\pagerange{\pageref{firstpage}--\pageref{lastpage}}
\pubyear{2005}

\begin{document}

\maketitle

\label{firstpage}

\begin{abstract}
{We present here new results of two-dimensional hydrodynamical simulations
of the eruptive events of the 1840s (the great) and the 1890s (the minor)
eruptions suffered by the massive star $\eta$ Car. The two bipolar nebulae
commonly known as the Homunculus and the little Homunculus were formed
from the interaction of these eruptive events with the underlying stellar
wind.  As in previous work (Gonzalez et al. 2004a, 2004b), we assume
here an  interacting, nonspherical multiple-phase wind scenario to
explain the shape and the kinematics of both Homunculi, but adopt a more
realistic parametrization of the phases of the wind. During the 1890s
eruptive event, the outflow speed {\it decreased} for a short period of
time. This fact suggests that the little Homunculus is formed when the
eruption ends, from the impact of the post-outburst $\eta$ Car wind
(that follows the 1890s event) with the eruptive flow (rather than by the
collision of the eruptive flow with the pre-outburst wind, as claimed
in previous models; Gonzalez et al. 2004a, 2004b). Our simulations
reproduce quite well the shape and the observed expansion speed of the
large Homunculus. The little Homunculus (which is embedded within the
large Homunculus) becomes Rayleigh-Taylor unstable and develop filamentary
structures that resembles the spatial features observed in the polar caps.
In addition, we find that the interior cavity between the two Homunculi
is partially filled by material that is expelled during the decades
following the great eruption. This result may be connected with the
observed double-shell structure in the polar lobes of the $\eta$ Car
nebula. Finally, as in previous work, we find the formation of tenuous,
equatorial, high-speed features that seem to be related to the
observed equatorial skirt of $\eta$ Car.}

\end{abstract}


\section{Introduction}

Located at a distance of 2.3 kpc, one of the most massive stars in
our Galaxy ever discovered, $\eta$ Car is a well-known example of
the evolved and unstable luminous blue variable (LBV) stars,
characterized by sporadic, violent mass-loss eruptive events (e.g.,
Humphreys $\&$ Davidson 1994). The 19th century spectrograms of
$\eta$ Car (see also Walborn $\&$ Liller 1977; Humphreys, Davidson
$\&$ Smith 1999 and references therein) provide evidence that the
star underwent a giant eruption during $\sim$20 yr around the 1840s,
where a few solar masses of gas ($\ge$10 M$_{\odot}$) were expelled
into the interstellar medium, and the total luminous output of
$\sim$10$^{49.5}$-$\sim$10$^{50}$ erg was comparable to a supernova
explosion (Smith et al. 2003a). It is still not understood how and
why this outburst occurred, but it may be connected with $\eta$ Car
being a binary star system (e.g., Damineli 1996; Damineli, Conti,
$\&$ Lopes 1997). From this event, a symmetric, bipolar nebula known
as the ``Homunculus" (that extends from -8 to +8 arcsec along its
major axis) was produced (Humphreys $\&$ Davidson 1994; Currie et
al. 1996; Davidson $\&$ Humphreys 1997; Smith $\&$ Gehrz 1998).
During the 1890s, the historical light curve of $\eta$ Car also
shows a much fainter eruptive event of shorter duration ($\sim$10
yr), but still an outburst in the sense that the mass loss rate was
enhanced compared with the normal underlying wind. This eruption
eruption resulted in the formation of a smaller nebula embedded
within the large Homunculus (with an angular size of about $\pm$2
arcsec) called the ``Little Homunculus" (LH) (Ishibashi et al. 2003;
Smith 2005). With a total mass of $\sim$0.1 M$_{\odot}$, the polar
caps of the LH currently move at slower speeds ($\sim$ 250 km
s$^{-1}$) than the large Homunculus in the polar direction
($\sim$650 km s$^{-1}$), but shares the same prolate geometry. The
total kinetic energy released in the 1890s event is $\sim$
10$^{46.5}$ erg (a factor of $\sim 10^3$ smaller than the kinetic
energy expelled during the great eruption). Apart from both
Homunculi, observations of $\eta$ Car with the Hubble Space
Telescope (e.g., Morse et al. 1998) also show the presence of an
equatorial skirt (orthogonal to the axis of the Homunculus) that
moves with velocities of 100-350 km s$^{-1}$ (Smith $\&$ Gehrz 1998;
Davidson et al. 2001), although high speed features, with typical
velocities of $\sim$750 km s$^{-1}$ (Smith et al. 2003b) or even
larger (Weis 2005) have also been detected. These equatorial
components may have been expelled from both the 1840s and the 1890s
outbursts (e.g., Davidson $\&$ Humphreys 1997). The large
Homunculus is the central portion of a larger nebula (the outer
ejecta) around $\eta$ Car that extends up to a diameter of
60 arcsec (e.g., Weis 2001, 2005; Walborn 1976; Walborn et al.
1978). The outer ejecta are not symmetric and contain numerous
filaments and knots that may have been produced by previous stellar wind
mass loss. The sizes and morphology of such structures in the outer
ejecta are manifold. Radial velocities for the knots reach up to
2000 km\,s$^{-1}$ (Weis 2001, Weis 2005), but the average expansion
velocity of the outer ejecta lies at lower values (around 750
km\,s$^{-1}$), similar to the speeds found in the Homunculus. The
outer ejecta also follow a bipolar pattern, like the large
Homunculus, with approximately the same axis of symmetry.

At optical wavelengths, the large Homunculus is mainly a
reflection nebula. Its spectrum shows the presence of dust scattered
emission (e.g., Hillier $\&$ Allen 1992) that allows one to see
indirectly the shape of the $\eta$ Car wind (see Smith et al. 2003b)
and also low-excitation intrinsic emission. In addition,
near-infrared spectra obtained by Smith (2006) confirmed the
existence of a double-shell structure at the edges of the
polar lobes of the Homunculus (which was previously inferred from
thermal dust emission; see Smith et al. 2003a). A thin outer shell
of intrinsic H$_{2}$ emission (that traces the main scattering layer
seen in visual images), and a thicker inner skin of [Fe II] (which
partially fills the interior of the lobes). On the other hand, the
outer ejecta are an emission nebula. Smith $\&$ Morse (2004) present
optical spectra of $\eta$ Car showing strong oxigen lines in some
emission features of the outer ejecta. At high energies, Chandra has
detected X-ray emission from the large-scale nebula around the star
(and also from the central object, probably due to a wind collision
region around the massive binary system; e.g., Corcoran et al. 2001;
Pittard $\&$ Corcoran 2002). Soft X-ray (0.1-0.8 keV) observations
by Seward et al. (2001) show an extended shell clearly associated
with the debris field of the outer ejecta which are compatible with
a collision process between a fast wind and cometary knots of
slower material. The $\eta$ Car nebula is also a bright radio source
consistent with thermal (free-free) emission. Observations by
Retallack (1983) at 1.415 GHz (taken with the Fleurs synthesis
telescope) show emission from an overall region as large as 40
arcsec (i.e., larger than the optical Homunculus). Gonz\'alez et
al. (2006) estimated the contribution by shocks to the total
radio-continuum emission detected from the $\eta$ Car nebula. Using
observational estimates of the wind parameters of the eruptive event
of the 1890s and of the stellar wind after the end of the eruption,
these authors investigated the evolution of the polar caps of the LH
formed as a result of the collision between these outflows. They found
that the LH emits continuum radiation which is detectable at radio
wavelengths and indeed, has an appreciable contribution to the total
flux of the $\eta$ Car nebula.

Different models have been proposed to explain the shaping and
kinematics of the $\eta$ Car bipolar nebulae (see, for instance,
Soker 2001; Soker 2004; Matt $\&$ Balick 2004; Gonzalez et
al. 2004a, b; A. Frank et al. 1995, 1998). One possible
explanation is that it is produced by the interaction of the  winds
expelled by the central star at different injection velocities
(e.g., Icke 1988; Frank, Balick $\&$ Davidson 1995; Dwarkadas $\&$
Balick 1998; Frank, Ryu $\&$ Davidson 1998; Langer, Garc\'ia-Segura
$\&$ Mac Low 1999; Gonz\'alez et al. 2004a,b). Adopting a colliding
wind scenario, Gonz\'alez et al. (2004a,b) performed two-dimensional
numerical simulations of the Homunculi nebulae of $\eta$ Car. In
their models, the large Homunculus is formed by the interaction of
the eruptive outflow of the 1840s with the pre-outburst $\eta$ Car
wind (both with different degrees of nonspherical symmetry). A second
eruption (assumed to be spherical) collides with the pre-outburst
wind giving rise to the LH. These authors showed that such a scenario could
explain the shape and kinematics of the Homunculi, and also the
existence of the high-velocity features observed in the equatorial
plane. However, their prescribed conditions to create the LH nebula
do not agree with overall results suggest by the observations. During
the 1890s event, the $\eta$ Car wind slowed down
and increased its mass loss rate, returning to its normal quiescent
state after the eruption ended. Another observational result (as
mentioned earlier) is that the Homunculus contains much more mass
than that that had been previously recognized (Smith et al. 2003a).
In this work, we perform new numerical simulations that incorporate
these observational results and investigate  the overall evolution
of the bipolar outflows of $\eta$ Car.

The paper is organized as follows. In $\S$ 2, we describe the model.
The numerical simulations and the discussion of the results are
presented in $\S$ 3, and in $\S$ 4 we draw our conclusions.

\section{The model}

In our model, we consider (see also Gonz\'alez et al. 2004a,b) a
simplified interacting wind scenario, in which a nonspherical outburst
collides with a slow wind also with asymmetric density and velocity 
distributions. We suppose that $\eta$ Car originated  a
nonspherical environment (prior to the major eruption of the 1840s)
from the ejection of a pre-outburst wind into a homogeneous ambient
medium. According to observational estimates, we also assume that
both the mass loss rate and the injection velocity were drastically
increased during the main eruption of the 1840s, after which the
original slow wind resumes. We also suppose that the wind parameters
are suddenly changed during the smaller mass-ejection event of the
1890s, after which the original wind again resumes. However,
observations of $\eta$ Car at this epoch (e.g. Whitney 1952; Walborn
$\&$ Liller 1977) give evidence that the expansion speed of the
stellar wind decreased during this eruption, rather than increasing
(see also Le Sueur 1870). Such variabilities in the wind parameters
of $\eta$ Car during the 19th century outbursts must have resulted
in the formation of a pair of double-shock wave structures (called
working surfaces; see Raga et al. 1990), that correspond - in our
colliding wind scenario - to the large and little Homunculi. Then,
the large Homunculus would be produced by the interaction between an
outburst wind (with a latitude-dependent velocity and density) and a
pre-eruptive slower wind (also with a nonspherical symmetry).
Nevertheless, the little Homunculus would be formed - unlike
previous numerical modeling (Gonz\'alez et al. 2004a,b) - from the
impact of the post-outburst wind with the outflow expelled during
the minor event, that is, when the eruption ended. According to
Gonz\'alez et al. (2006), a sudden increase in the ejection velocity
at the end of the eruption instantaneously occurs and forms
the inner nebula at the base of the wind (as the fast upstream
flow begins to be ejected).

\begin{figure}
\includegraphics[width=84mm]{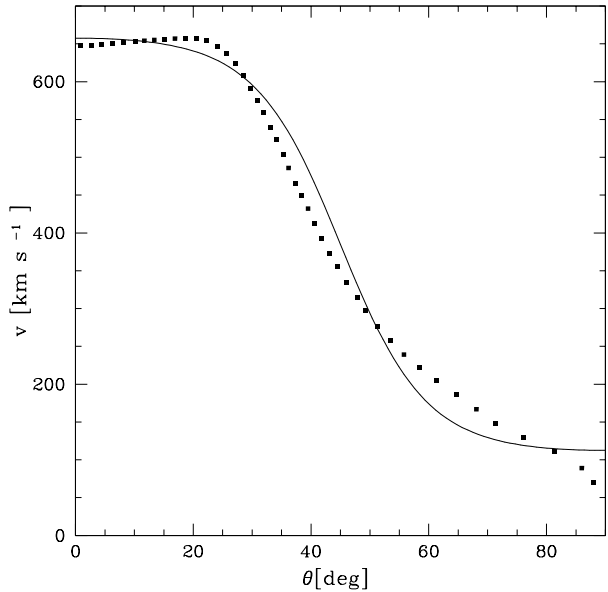}
 \caption{The solid line gives the best fitting curve to the
measured velocity in H$_2$ (squares) by Smith (2006) at different
latitudes of the $\eta$ Car Homunculus (with $\theta$= 0$^{\circ}$
at the pole). For this fitting we have assumed $\lambda=$ 1.9,
$v_{1}$= 670 km s$^{-1}$, $v_{2}=$ 100 km s$^{-1}$, which gives (see
eq. [\ref{homvelobs}]) a polar expansion velocity $v_p$= 657.53 km
s$^{-1}$ and $v_e$= 112.47 km s$^{-1}$ at equator.}
\end{figure}

In order to estimate the flow parameters (such as the injection
velocity $v$ and the number density $n$) for the different wind
phases, we have fitted a curve to the expansion speed of the outer
H$_2$ shell of the $\eta$ Car Homunculus measured by Smith (2006).
From this fitting, we find a latitude-dependent velocity
given by,

\begin{eqnarray}
v= v_{1}\,F(\theta),
\label{homvelobs}
\end{eqnarray}

\noindent
with
\begin{eqnarray}
F(\theta)= {{(v_{2}/v_{1}) + e^{2\,z}}\over{1 + e^{2\,z}}},
\nonumber
\end{eqnarray}

\begin{figure}
\includegraphics[width=84mm]{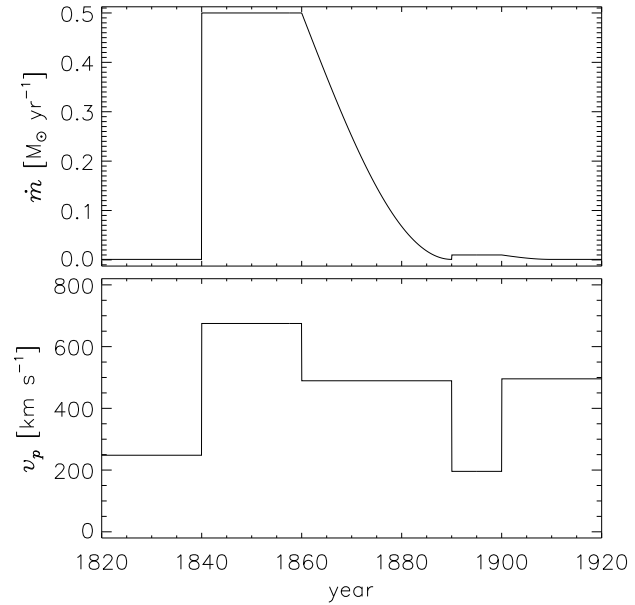}
 \caption{ Behaviour of the adopted parameters of the
 interacting outflows. We show the mass loss rate (top panel)
 and the wind velocity at the poles (bottom panel) as functions
 of time. During the great eruption of the 1840s ($\sim$20 yr), both
 parameters were suddenly increased, while during the minor event of
 the 1890s the flow velocity decreased for $\sim$10 yr. In addition, we
 assume a gradual fading in the mass loss rate during the years after
 the end of both events, asymptotically approaching the mass loss
 rate of the underlying wind (10$^{-3}$ M$_{\odot}$ yr$^{-1}$). }
\end{figure}

\noindent where the parameter $z=\lambda\, \mbox{cos}(2\theta)$
controls the shape of the Homunculus, $\lambda$ is a constant and
$\theta$ is the polar angle. The velocities $v_{1}$ and $v_{2}$ are
related to the expansion speed in the polar ($\theta$= 0$^{\circ}$)
and equatorial ($\theta$= 90$^{\circ}$) directions. In Figure 1, we
present the best fit to  Smith (2006)'s observations obtained with
$\lambda=$ 1.9, $v_{1}$= 670 km s$^{-1}$, $v_{2}$= 100 km s$^{-1}$.
With these parameters, the predicted expansion velocity in the polar
and equatorial directions are $v_p$= 657.53 km s$^{-1}$ and $v_e$=
112.47 km s$^{-1}$, respectively. Using this fit of the expansion
velocity of the outer Homunculus, we estimate the flow parameters of
the interacting winds as follows.

\begin{table*}
\begin{center}
TABLE 1
\vskip 0.5cm
\centerline{\sc Parameters of the colliding outflows}
\vskip 0.2cm
\begin{tabular}{ccccccccccccc}
\hline
\hline
\noalign{\smallskip}
Wind Phase & & $\lambda$ & & $v_1$ [km s$^{-1}$] & & $v_2$ [km s$^{-1}$] & &
$v_p$ [km s$^{-1}$] & & $v_e$ [km s$^{-1}$]
 & & $\dot m$ [M$_{\odot}$ yr$^{-1}$]\\
 & & & & & & & & $(\theta= 0^{\circ})$ & &$(\theta= 90^{\circ})$ & & \\
\hline
{\it Pre-outburst wind} & & 2.4 & & 250.00 & & 14.00 & &
 248.07 & & 15.93 & & 10$^{-3}$ \\
{\it Great Eruption} & & 1.9 & & 687.76 & & 102.65 & &
 674.95 & & 115.45 & & 5 $\times$ 10$^{-1}$ \\
{\it Post-outburst wind} & & 1.9 & & 500.00 & & 14.00 & &
 489.37 & & 24.63 & & $10^{-3}\times \phi(t)$ $\, ^{(\dagger)}$ \\
{\it Minor Eruption} & & 1.9 & & 200.00 & & 10.00 & &
 195.84 & & 14.16 & & 10$^{-2}$ \\
{\it Post-outburst wind} & & 1.9 & & 500.00 & & 300.00 & &
 495.62 & & 304.38 & & $10^{-3}\times \varphi(t)$ $\, ^{(\ddagger)}$ \\
\noalign{\smallskip}
\hline
\end{tabular}

\begin{flushleft}
\hskip 1.5cm {($\dagger$) see eq. [\ref{phi}]}\\
\hskip 1.5cm {($\ddagger$) see eq. [\ref{varphi}]}\\
\end{flushleft}
\end{center}
\end{table*}

From a straightforward application of the formalism developed by
Cant\'o et al. (2000) for outflows with time-dependent injection
velocities (see also Gonz\'alez $\&$ Cant\'o 2002; Cant\'o et al.
2005), we studied the dynamical evolution of the Homunculus. Let us
first consider the downstream wind with latitude-dependent injection
velocity $v_0$ (given by eq. [\ref{homvelobs}]) and constant mass
loss rate per unit solid angle. Then, the injection number density
of the wind is equal to,

\begin{eqnarray}
n= n_0\,{\biggl({{r_0}\over{r}}\biggr)}^2
         {{1}\over{F(\theta)}},
\label{dens}
\end{eqnarray}

\noindent
where $n_0 (= \dot m_0/4\pi \mu v_0 r^2_0;$ being $\mu$ the
mean mass per nucleon and $\dot m_0$ the total mass loss rate)
is the number density at the injection
radius $r_0$ (assumed to be a few stellar radii). Note that
$F(\theta)$ depends on the particular values of $v_1$,
$v_2$, and $z$ of the pre-outburst wind.

When the great eruption begins, the wind parameters are suddenly
increased to $a v_0$ and $b \dot m_0$, respectively (where $a$ and
$b$ are constants). Such a variation in the flow parameters forms
instantaneously (at the base of the wind) a working surface (which
will correspond in our model to the large Homunculus) that
moves with a constant velocity given by,

\begin{eqnarray}
v_{ws}= \sigma v_0,
\label{homvel}
\end{eqnarray}

\noindent
with $\sigma= (a^{1/2} + a\,b^{1/2})/ (a^{1/2} + b^{1/2})$.
This velocity is intermediate between the low-velocity
downstream wind ($v_0$) and the faster upstream outflow
($a\,v_0$).

Given (for instance) $v_0$ and $v_{ws}$, equation (\ref{homvel})
gives $\sigma$ as a function of $a$ and $b$. We estimate $b$
from observations of the mass contained in the Homunculus and the
duration of the major eruption (and compare with the mass
loss rate of the standard wind of $\eta$ Car) and then, we obtain
$a$. Assuming the same latitude-dependent velocity and density (eqs.
[\ref{homvelobs}] and [\ref{dens}], respectively) for the
pre-outburst wind and the great eruption, and using the fit to Smith
(2006)'s observations of the expansion velocity of the $\eta$ Car
Homunculus (Fig. 1), we finally obtain the flow parameters at the
different phases of the interacting winds, keeping $v_0$ as a free
parameter.

The historical light curve of $\eta$ Car shows a sudden increase in
brightness when the eruptions of the 1840s and 1890s turned-on and a
gradual fading during the years after the end of the eruptive events (e.g.
Walborn $\&$ Liller 1977; Davidson 1987; Humphreys $\&$ Davidson
1994; Humphreys, Davidson $\&$ Smith 1999). This suggests that the
post-outburst mass-loss rates of the major ($\dot m_{1}$) and the minor
($\dot m_{2}$) events  decrease, approaching the mass loss rate
of the underlying wind ($\dot m_0= 10^{-3}$ M$_{\odot}$ yr$^{-1}$;
e.g. Humphreys $\&$ Davidson 1994; Davidson $\&$ Humphreys 1997).
We then consider,

\begin{eqnarray}
\dot m_{1} = \dot m_0 \, \phi(t),
\label{mpe1}
\end{eqnarray}

\noindent
and

\begin{eqnarray}
\dot m_{2} = \dot m_0 \, \varphi(t),
\label{mpe2}
\end{eqnarray}

\noindent
respectively, with

\begin{eqnarray}
\phi(t)= {{\dot m_{ge}}\over{\dot m_0}} +
     \biggl(1- {{\dot m_{ge}}\over{\dot m_0}} \biggr)
     \,\mbox{sin} \biggl[{{\pi}\over{2}}
     \biggl({{t-t_{1}}\over{\Delta t_1}}
     \biggr) \biggr] \, ,
\label{phi}
\end{eqnarray}

\noindent
and,

\begin{eqnarray}
\varphi(t)= {{\dot m_{me}}\over{\dot m_0}} +
     \biggl(1- {{\dot m_{me}}\over{\dot m_0}} \biggr)
     \,\mbox{sin} \biggl[{{\pi}\over{2}}
     \biggl({{t-t_{2}}\over{\Delta t_2}}
     \biggr) \biggr] \, ,
\label{varphi}
\end{eqnarray}

\noindent
where $\dot m_{ge} (= 5 \times 10^{-1}$ M$_{\odot}$
yr$^{-1}$) and $\dot m_{me} (= 10^{-2}$ M$_{\odot}$ yr$^{-1}$) are
the estimated mass loss rates during the eruptions, $t_1$ and $t_2$
correspond to the transition times when both events end, and $\Delta
t_1 (= 30$ yr) and $\Delta t_2 (= 10$ yr) represent in our model the
duration of the post-eruption phases. In Figure 2, we show
the behaviour of the adopted parameters (the mass loss rate and the
ejection velocity at the poles) of the interacting outflows as
functions of time.

\section{Numerical Simulations}

We have perfomed gasdynamic 2D numerical simulations (considering
axial symmetry) of the 19th century outbursts of $\eta$ Car.
Adopting the colliding wind model described in $\S$2, we use the
adaptive-grid YGUAZ\'U-A code originally developed by Raga et al.
(2000; see also Raga et al. 2002) and modified by Gonz\'alez et al.
(2004a, b). This code integrates the hydrodynamic equations
explicitly accounting for the radiative cooling with a set of
continuity equations for the atomic/ionic species HI, HII, HeI,
HeII, HeIII, CII, CIII, CIV, OI, OII, and OIII. The flux-vector
splitting algorithm of Van Leer (1982) is employed. The simulations
were computed on a five-level binary adaptive grid with a maximum
resolution of 3.9 $\times$ 10$^{14}$ cm, corresponding to 1024
$\times$ 1024 grid points extending over a computational domain of
($4 \times 10^{17}$ cm) $\times$ ($4 \times 10^{17}$ cm). The
adopted abundances (by number) for the different elements are (H,
He, C, O) = (0.9, 0.099, 0.0003, 0.0007).

\begin{figure*}
\includegraphics[width=174mm]{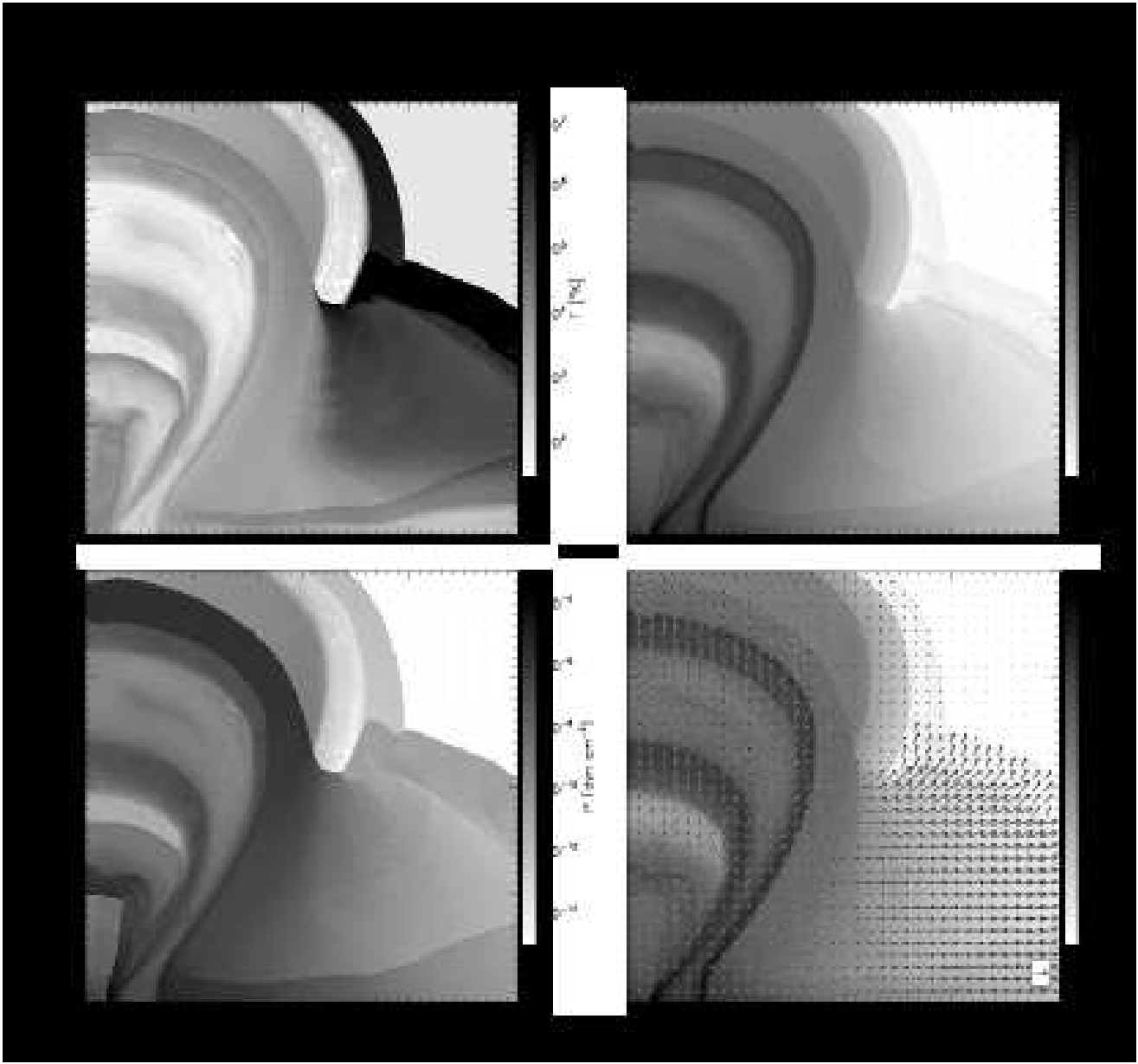}
 \caption{Stratifications of temperature (top left), density
(top right), pressure (bottom left), and velocity-field superposed
on the density map (bottom right) obtained 169 yr after the great
eruption of $\eta$ Car (i.e., around  2009) are presented. The arrow
shown at the lower right corner of the velocity-field stratification
corresponds to 1000 km s$^{-1}$ (see the text for more details)}.
\end{figure*}

\subsection{Initial Physical conditions}

In Table 1 (see also Fig. 2), we list the adopted parameters
for the interacting winds. We have assumed different equator-to-pole
velocity (and density) contrasts.

Initially, the computational domain is filled by a
homogeneous ambient medium with temperature $T_a=$ 10$^2$ K and
density $n_a=$ 10$^{-3}$ cm$^{-3}$. Afterwards, a gaseous toroidal
distribution is formed by the injection of a pre-outburst wind into
this unperturbed environment (with a different  $\lambda$ parameter
from the other injected later wind phases which controls
the shape of the wind). For this wind, we have assumed a terminal
velocity (in the polar direction)  $v_1=$  248.07 km s$^{-1}$
(Gonz\'alez et al. 2004a,b) and a mass loss rate of 10$^{-3}$
M$_{\odot}$ yr$^{-1}$ which was injected at a distance of
$r_0$=10$^{16}$cm (from the stellar surface) with a temperature of
$T_0$=10$^{4}$ K. Once a poloidal environment is produced,
a much faster (= 674.95 km s$^{-1}$) and massive outflow (5
$\times 10^{-1}$ M$_{\odot}$ yr$^{-1}$) is expelled during the
estimated duration of the great eruption ($\sim$ 20 yr; e.g.
Davidson $\&$ Humphreys 1997). For this  outburst phase
(that dominates the momentum flux) we have adopted the same degree
of nonspherical symmetry as the observed for the large Homunculus,
as given in Figure 1. After this event, a third outflow (the
post-outburst wind) is turned on with similar conditions to the
current stellar wind of $\eta$ Car (e.g. Le Sueur 1870; Smith et al.
2003b), that is, we assume an injection velocity of
489.37 km s$^{-1}$ at the poles and a time-dependent mass loss rate
that approaches 10$^{-3}$ M$_{\odot}$ yr$^{-1}$ (eqs. [\ref{mpe1}]
and [\ref{phi}]). In order to account for the minor eruption, during
10 years (and 50 years after the great eruption event), we assumed
a slower eruptive wind with a terminal velocity - along the symmetry axis -
of $\sim$ 195.84 km s$^{-1}$ and a mass loss rate of 10$^{-2}$
M$_{\odot}$ yr$^{-1}$. After this outflow, a faster (= 495.62
km s$^{-1}$), but less massive wind with a mass loss rate approaching
to 10$^{-3}$ M$_{\odot}$ yr$^{-1}$ (eqs. [\ref{mpe2}] and [\ref{varphi}])
resumes. We also have assumed for this wind a higher speed (= 304.38
km s$^{-1}$) at equator, which is consistent with the present-day
latitudinal structure in $\eta$ Car's stellar wind (Smith et al. 2003b).

We should notice that Smith (2006) calculated the mass loss
distribution assuming uniform density as a function of latitude with
a constant width. These are not the assumptions of our current models.
Instead, we assumed that both the ejection velocity and the injection
density depend on latitude (eqs. [1] and [2], respectively) and therefore,
a constant mass loss rate per unit solid angle is adopted (see $\S$2).
However, a similar scenario to the Smith's findings was previously study
in Gonzalez et al. (2004b), where we assumed a nonspherical outburst wind
of the 1840s impinging on a slow pre-outburst wind with a larger mass-loss
rate in the polar direction (run D). We found that this scenario does not
develop significant equatorial features.

\subsection{Results of the simulations}

In this section we present the two-dimensional hydrodynamical
numerical simulations performed for the outbursts of $\eta$ Car,
adopting the colliding wind scenario described in $\S$ 2. Figure 3
shows the temperature, density, pressure, and velocity maps computed
for the interaction of the five winds above at a time $t$= 169 yr of
evolution after the great eruption. As  predicted in $\S$2,
the simulations show the formation of the outer Homunculus with a
double-shock structure, having an inner shock that decelerates the
fast outburst flow and an outer shock that accelerates the lower
velocity precursor wind. At the poles, a cold thin
shell behind the inward shock, and a hotter and thicker region
behind the outward shock are formed. This difference is due to the
radiative cooling of the shocked material behind both shocks (see
Gonz\'alez et al. 2004a). We find that for the great eruption, the
momentum flux is dominated by the material expelled during the
eruptive event (that is, the pre-outburst wind to the great
eruption momentum ratio, $\dot m_0\, v_0$/ $\dot m_{ge}\, v_{ge}$
$\ll$ 1 (where $\dot m_{ge}$ and $v_{ge}$ are the mass loss rate and the
wind velocity during the eruption, respectively), so that the
pre-outburst wind has a minor effect on the kinematics and
morphology of the bipolar lobes. Since the mass loss rate during the
eruption is more than two orders of magnitude (a factor of 500; see
Table 1) larger than that of the pre-outburst wind, the Homunculus
retains almost the same degree of asymmetry imprinted in the 1840s
event near the star.

\begin{figure}
  \includegraphics[width=1.\hsize]{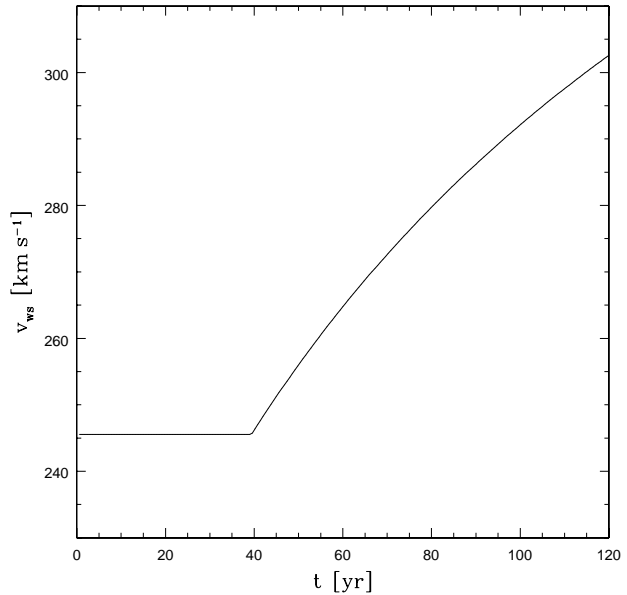}
  \caption{Behaviour of the velocity of the polar caps of
the inner Homunculus as a function of time. Initially they move with
constant velocity ($\sim$ 245.55 km s$^{-1}$). At a critical time
$t$= 39.4 yr (when the downstream shock disappears), the caps begin
to be accelerated, asymptotically approaching the velocity of the
post-outburst wind ($\sim$ 495.62 km s$^{-1}$). After 109 yr of
evolution (i.e., around the  year of 2009) they expand at $\sim$ 297
km s$^{-1}$. This value is consistent with the observed current
expansion speed of $\sim$ 250 km s$^{-1}$ of the little Homunculus
(Smith 2005).}
  \label{fig4.eps}
\end{figure}

We note that the shape and kinematics of the outer expanding
shell resembles that of the large Homunculus previously simulated by
Gonz\'alez et al. (2004a, b), but important differences from these
previous models are identified in the embedded structures of Figure
3. In the case of the minor eruptive event, the momentum
flux is also dominated by the eruption, but the post-outburst wind
has a significant effect on the kinematics and morphology of the
inner Homunculus once the low-velocity downstream material
is completely incorporated to the layer. Gonz\'alez et al.
(2006) have shown that this happens at a critical time $t_c= (\Delta
t)/(\sigma-1)$, being $\Delta t$ (=10 yr) the duration of the
eruptive event. Adopting the flow parameters given in $\S$ 3.1
($\sigma \simeq$ 1.25; a= 2.53 and b=0.1), we obtain $t_c\simeq$
39.4 yr. In Figure \ref{fig4.eps}, we show the behaviour of the
velocity of the polar caps of the inner Homunculus as a function of
time. Initially, they move with constant velocity ($\sim$ 245.55 km s$^{-1}$;
see eq. [3]), until the downstream shock disappears and a one-shock
structure stage begins. Later, the polar caps begin to be
accelerated, asymptotically approaching the velocity of the
post-outburst wind (see also Gonz\'alez $\&$ Cant\'o 2002). After
109 yr of evolution (around year 2009), they expand at $\sim$ 297 km
s$^{-1}$ and are located at a position $r_s\, \simeq$ 9.07$\times$
10$^{16}$ cm from the star. At a distance of 2.3 kpc, $r_s$
corresponds to an angular size of $\pm$ 2.6 arcsec, which is
consistent with the angular extent of roughly $\pm$2 arcsec (along
the major axis) of the inner Homunculus measured by Ishibashi et al.
(2003). This is also consistent with the angular size ($\pm$ 3 arcsec)
observed in the simulations.

\begin{figure}
  \includegraphics[width=1.\hsize]{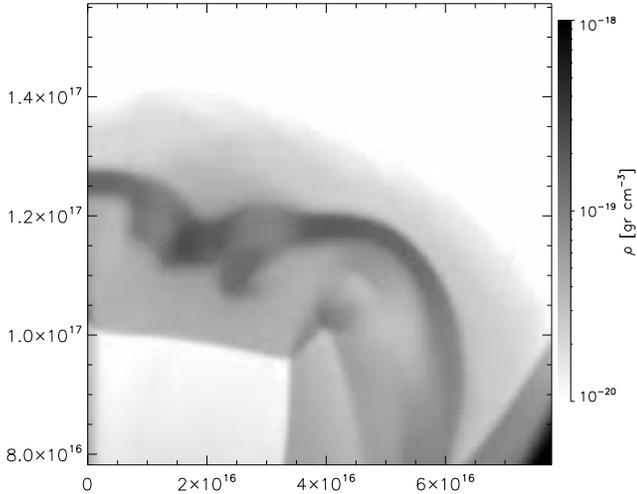}
 \caption{Detail of the density map of Fig. 3, showing the little
Homunculus at $t$= 119 yr after the minor eruptive event of $\eta$ Car
(or around the year 2009). The low-density fast wind (post-outburst wind)
pushing the high-density slow wind (minor eruption) causes the polar caps
of the little Homunculus to become  Rayleigh-Taylor unstable.}
  \label{fig5.eps}
\end{figure}

The inner Homunculus quickly becomes Rayleigh-Taylor
unstable (see figure \ref{fig5.eps}) due to the interaction of the
low-density fast wind (post-outburst wind) that pushes and accelerates
the high-density slow wind (minor eruption). The growth
time for this instability can be estimated as follows.

The density of the fast wind at the position of the shock, $r_{s}$,
is obtained by,

\begin{equation}
  \rho_{fw} = {{\dot M_{fw}}\over{4 \pi v_{fw} r_{s}^2}},
\end{equation}

\noindent where $\dot M_{fw}$ is the mass loss rate of the wind, and
$v_{fw}$ is its velocity at $r_s$. On the other hand, the
shock-bounded layer of the slow wind has a surface density
$\sigma_{sw}$ given by,

\begin{equation}
  \sigma_{sw} = {{M_{sw}}\over{4 \pi r_s^2}},
\end{equation}

\noindent where $M_{sw} = \dot M_{sw}\,\Delta t$ is the mass in the
shock-bounded layer, $\Delta t$ is the  time interval of  injection
of the slow wind, and $\dot M_{sw}$ is its mass loss rate. In the
above, we have assumed that the shock-bounded layer is thin.

Initially, the shock-bounded layer (and the slow wind shock) moves
with a velocity $v_s$ given by (eq. [\ref{homvel}]),

\begin{equation}
  v_s= \sigma v_{sw}
\end{equation}

\noindent where $v_{sw}$ is the velocity of the slow wind. In the
shock reference frame, the fast wind moves with velocity $v = v_{fw}
- v_s$, and thus excerts a hydrodynamical pressure $\rho_{fw} v^2$
on the shock-bounded layer. This layer will therefore, experience an
acceleration $a_s = \rho_{fw} v^2 / \sigma_{sw}$. Taking the values
of the mass loss rates and wind velocities from Table 1, we obtain a
value for the acceleration experienced by the layer of a$_s$ = 3.96
$\times 10^{-3}$ cm s$^{-2}$.

The dispersion relation for the Rayleigh-Taylor (RT) instability is
calculated by,

\begin{equation}
  \omega_k = -{{\rho_2-\rho_1}\over{\rho_2+\rho_1}} g k,
\end{equation}

\noindent where $\rho_2 > \rho_1$ are the densities of the fluid
layers, $g$ is the  acceleration, and $k$ is a wave number. We
identify $\rho_2 = \rho_{sw}$, $\rho_1 = \rho_{fw}$ and $g=a_s$.
Considering that the growth rate for the RT instability is largest
for the shortest wavelengths, we adopt the minimum resolved
wavelength as 3 times the resolution of our simulation,
$\lambda_{RT} = 1.17 \times 10^{15}$ cm. With $\rho_{fw}= 10^{-18}$
g cm$^{-3}$ and $\rho_{sw} = 2 \times 10^{-17}$ g cm$^{-3}$, the
RT instability grows in about one year, in agreement with the
simulations.

In Figure 6, we show that our model of interacting nonspherical
winds can explain the formation of the large and little Homunculi.
From the simulations, we have found an appropriate combination of
the flow parameters (which control the degree of asymmetry) that
best matches the kinematic structure of the thin outer shell of
H$_2$ observed by Smith (2006). In contrast with our previous models
of $\eta$ Car (Gonz\'alez et al. 2004a,b) in which we observed a hollow
nebula, it is noteworthy that the new numerical experiments show a
thicker layer (with a density between $\simeq 10^{-20}$ g cm$^{-3}$
and $10^{-18}$ g cm$^{-3}$ along the symmetry axis) partially filling the
interior of the lobes of the large Homunculus. It may be 
related with the inner skin of [Fe II] emission detected
by Smith (2006).

\begin{figure}
  \includegraphics[width=1.\hsize]{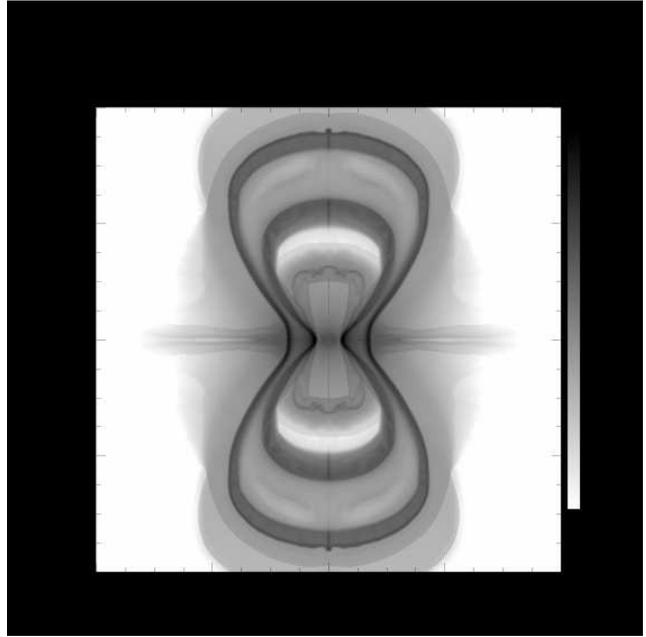}
 \caption{Present-day structure of the $\eta$ Car nebulae. The inner
and outer Homunculus and a tenuous equatorial skirt are depicted. A
thick dense layer between both Homunculi is also observed.}
  \label{fig6.eps}
\end{figure}

On the other hand, as in the previous work, the high-velocity
ejections at equator, which arise from the impact of the external
Homunculus on the shock front of the pre-eruptive wind, resemble the
observed equatorial skirt of $\eta$ Car. At a time $t\simeq 100$ yr
after the great eruption turn-on, the impact at low latitudes occurs
producing a hot ($T\simeq 10^7$ K), tenuous ($\rho\simeq 10^{-18}$
g cm$^{-3}$) structure. This structure is accelerated, reaching an
expansion velocity of $\sim$1000 km s$^{-1}$ and produces fast ejected
material that could be related to the observed material confined to
the equatorial plane of $\eta$ Car. The gas is concentrated to
the equator by the impact of the two shock fronts, and then it flows
into an environment of decaying pressure. Why are these equatorial
features moving faster and what accelerate them in our simulations
are still open issues which we will attend in a forthcoming paper.
This effect has been also addressed in Gonz\'alez et al. (2004a,b)
and it is probably connected with a {\it nozzle} problem.  As can be seen
in the simulations (Figs. 3 and 6), the equatorial ejection has some
resemblance with a supersonic flow through a tube of varying
cross-sectional area (the problem of the {\it de Laval nozzle}). At
supersonic speeds, an increase in velocity is expected when the area
of the nozzle increases, such as our models show. \footnote{We notice
that the effect described her is distinct from
the one discussed, e.g., in Akashi $\&$ Soker (2008). These authors
claim the presence of jets (or collimated fast winds) launched by the
by the central star (or by a companion) to explain the presence of an
expanding disk and then relate its acceleration to compression of the
material in the equatorial plane by the two lobes.}

\section{Discussion and Conclusions}

In this article, we carried out high-resolution two-dimensional
gasdynamic simulations of the dramatic outbursts suffered by the
star $\eta$ Car during the 1800s, the larger of which occurred in
the 1840s and resulted in the formation of the large Homunculus
nebula and the smaller one in the 1890s created the inner little
Homunculus. During these events, the parameters of the $\eta$ Car
wind may have drastically changed in short periods of time. In
contrast with the great eruption (where both the mass loss rate and
the ejection velocity were suddenly increased), during the minor
event of the 1890s (much fainter and of shorter duration than the
1840s event) the mass loss rate was enhanced, but the wind velocity
diminished compared with the normal $\eta$ Car wind.

Considering a simplified interacting stellar wind scenario, we could
explain the shape and the observed expansion speed (as a function of
latitude) of the large Homunculus. In addition, our numerical models
show that the little Homunculus is formed at the $end$ of the 1890s
eruption, when the post-outburst $\eta$ Car wind collides with the
eruptive outflow. Important differences with regard to our previous
models of the $\eta$ Car nebula (Gonz\'alez et al. 2004a,b) have been
obtained. In Gonz\'alez et al. (2004a,b), we assumed that
the impact between the 1890 outburst and the pre-outburst
wind was the cause of the development of the little Homunculus nebula.
As a matter of fact, although the momentum flux is dominated
by the 1890s eruption, the post-outburst wind has an important
effect on the kinematics and morphology of the little Homunculus as
it causes  the complete deposit of the material expelled during the
eruption into the polar caps. At this time, the  powerful
stellar wind accelerates the inner Homunculus material
so that it asymptotically reaches the velocity of the post-outburst
wind. Due to  Rayleigh-Taylor instabilities  generated by the
low-density fast wind pushing the high-density slow wind, the polar
caps of the inner Homunculus quickly develop filamentary structuring
(Fig. 5) that shows some resemblance with the observed spatial
structures in the polar lobes of the little Homunculus by Smith
(2005) that suggest that it is not perfectly homologous
to the larger Homunculus nebula.

Also, in contrast with the previous models of $\eta$ Car
(Gonz\'alez et al. 2004a,b), it is noteworthy that in the present
numerical modeling the interior cavity between the Homunculi is
partially filled by material that is expelled during the decades
following the end of the 1840s great eruption, rather than being
almost empty. This agrees with the observed double-shell structure
observed in the polar lobes by Smith (2006), consisting of a thin
outer H$_2$ skin (which contains most of the material of
the nebula) and a thicker [Fe II] layer with a more irregular spatial
distribution.

As in the previous models (Gonz\'alez et al.
2004a,b), the present results show the formation of an
equatorial outflow with both low and high velocity features. These
are probably related to the equatorial skirt of $\eta$ Car. We note
however, although it has been predicted that the observed skirt
might be associated with the two outbursts (e.g. Davidson et al. 2001),
our simulations indicate that only the great eruption contributes to its
formation.

A final remark is in order. One could ask whether one can learn
about the progenitor of Eta Car from the assumed mass loss
history or what can cause such sharp changes in the mass loss
rate and speed, without affecting the general shape of the Homunculi.
As stressed in Gonz\'alez et al. (2004a,b). these questions are
vey probably connected with Eta Car being a binary star system
and the nature of the interaction between the main and companion
stars. In this work, however, we have focussed in the formation
and dynamical evolution of the shock structures associated to the
large and little Homunculi given the parameters (terminal velocity,
mass-loss rate and/or density) of the different wind phases,
without addressing the inner mechanism that first triggered it or
its variability at the much smaller scales of the central source of
Eta Car. The orbit of the secondary star, for instance, has
apoastro and periastro distances of $\sim$14 AU and $\sim$3 AU, respectively,
while in our simulations, the outflows are injected at much greater
distances (10$^{16}$ cm). A detail study univocally connecting both
the inner source scales and the outer scales was, therefore, out of
the scope of the present study. Nonetheless, there have been some
first efforts in this direction (e.g., Soker 2001 ; Falceta-Gon\c{c}alvez
$\&$ Abraham 2009) that should be further explored in the future.

\section*{Acknowledgments}

The work of RFG and AVC was supported by the DGAPA (UNAM)
grant IN 117708. GCG thanks DGAPA (UNAM) grant
IN 106809 and CONACyT grant 50402-F. AR, JC, and PFV acknowledge
financial support by CONACyT grants 46828-F and 61547. EMGDP has
been partially supported by the Brazilian Foundations FAPESP
(2006/50654-3) and CNPq grants. The authors have benefited from
elucidating conversations and comments from Z. Abraham and
M.A. de Avillez. The authors also thank the useful comments
of the referee Noam Soker.

\end{document}